\documentclass[prd,aps,floats,twocolumn,10pt]{revtex4}
\usepackage{epsfig}

\newcommand{\bc}{\begin{comment}}
\newcommand{\ec}{\end{comment}}

\begin{document}

%\draft
\preprint{\
\begin{tabular}{rr}
%CfPA/96-th-15 &  \\
&
\end{tabular}
}
%\twocolumn[\hsize\textwidth\columnwidth\hsize\csname@twocolumnfalse\endcsname
\title{Constraining Lorentz violation with cosmology}
\author{J.A.~Zuntz$^{1}$, P.G.~Ferreira$^{1}$ and T.G~Zlosnik$^{1}$}
%
%\address{
\affiliation{
$^1$Astrophysics, University of Oxford, Denys Wilkinson Building, Keble Road, Oxford OX1 3RH, UK\\\
}

\begin{abstract}
The Einstein-Aether theory provides a simple, dynamical mechanism for breaking Lorentz invariance. 
It does so within a generally covariant context and may emerge from quantum effects in more fundamental theories. The theory leads to a preferred frame and can have distinct experimental signatures. In this letter, we perform a comprehensive study of
the cosmological effects of the Einstein-Aether theory and use observational data to constrain it.  Allied to previously determined consistency and experimental constraints, we find that an Einstein-Aether universe can fit experimental data over a wide range 
of its parameter space, but requires a specific rescaling of the other cosmological densities.

\end{abstract}

\date{\today}
\pacs{PACS Numbers : }
%]
\maketitle

%%%%%%%%%%%%%%%%%%%%%%%%%%%%%%%%%%%%%%%%%%%%%%%%%%%%%%%%%%%%%%%%%%%%%
\section{Introduction}

The spacetime symmetry of local Lorentz invariance is a cornerstone of modern physics \cite{weinberg}, but is not inviolate.  Violations  can occur in quantum gravity theories, with the symmetry emergent and approximate at macroscopic levels\cite{QG}.  In the particle physics sector the symmetry has been experimentally verified to extremely high precision \cite{kostelecky}.  On the large scales characteristic of the gravitational sector, however, constraints are much less certain.  In this letter we explore the extent to which precision cosmology can constrain a Lorentz-violating theory.

The theoretical workhorse for studying violation of Lorentz symmetry in gravitation is the Einstein-Aether theory \cite{jacobson1}, a simple, elegant proposal for dynamically violating Lorentz invariance within
the framework of a diffeomorphism-invariant theory. It is a refinement of the gravitationally coupled vector field theories 
first proposed by Will and Nordvedt in 1972 \cite{WN} and has been explored in exquisite
detail by Jacobson, Mattingly, Foster and collaborators \cite{jacobson2, mattingley, foster}. 
A Lorentz-violating vector field, henceforth called the {\it aether}, will affect cosmology: 
it can lead to a renormalization of the Newton constant \cite{CL}, leave
an imprint on perturbations in the early universe \cite{lim,kanno}, and in more elaborate actions it may even affect the
growth rate of stucture \cite{ZFS2,H1}. Preliminary calculations have been done on the affect of the aether on anisotropies
of the cosmic microwave background \cite{mota}.

The action for the Einstein-Aether is:
\begin{eqnarray*}
S=\int d^4x \sqrt{-g}\left[\frac{1}{16\pi G}R+{\cal L}(g^{ab},A^{b})\right]
+S_{M}
\end{eqnarray*}
where $g_{ab}$ is the metric, $R$ is the Ricci scalar of that metric,
$S_M$ is the matter action, and $\cal{L}$ is constructed to be generally
covariant and local.  $G$ is the bare gravitational constant, not necessarily equal to the locally measured value.  $S_M$ couples only to the metric $g_{ab}$ and not to 
$A^{b}$ and
\begin{eqnarray*}
{\cal L}(g^{ab},A^{b})\equiv\frac{1}{16\pi G}[K^{ab}_{\phantom{ab}cd} \nonumber
\nabla_a A^{c}\nabla_b A^{d}
+\lambda(A^b A_b+1)]  ,
\end{eqnarray*}
where $K^{ab}_{\phantom{ab}cd}\equiv c_1g^{ab}g_{cd}
+c_2\delta^{a}_{\phantom{a}c}\delta^{b}_{\phantom{b}d}+
c_3\delta^{a}_{\phantom{a}d}\delta^{b}_{\phantom{b}c}
-c_{4}A^{a}A^{b}g_{cd}$ \cite{conventions}.  We will use the notation $c_{12...}\equiv c_{1}+c_{2}+..$.
The gravitational field equations for this model take the form:
$G_{ab}=\tilde{T}_{ab}+8\pi GT_{ab}$
where the stress-energy tensor for the vector field $\tilde{T}_{ab}$ is given in \cite{jacobson1} and $T_{ab}$ describes the conventional fluids.

A number of constraints on the $c_i$s have been derived. Most notably
a Parametrized Post Newtonian (PPN) analysis of the theory leads to a
reduction in the dimensionality of parameter space such that $c_2$ and $c_4$ can be
expressed in terms of the other two parameters: $c_2=(-2c_1^2-c_1c_3+c_3^2)/3c_1$ and
$c_4=-c_3^2/c_1$ \cite{foster}.  
Additionally, the squared speeds of the gravitational and aether waves with respect to the preferred frame must be greater than one
so as to prevent the generation of vacuum \v{C}erenkov radiation by cosmic rays \cite{cerenkov}.  We shall
 label this space of models as ${\cal C}$.  A final constraint arises from considering the effects of the aether on the 
damping rate of binary 
pulsars. The rate of energy loss in such systems by gravitational radiation agrees with the prediction of General Relativity to one part in $10^{3}$. It has been shown 
\cite{bfosterweak} that, for the Einstein-Aether theory to agree with General Relativity for these systems, we require that $c_+\equiv c_1+c_3$ and  $c_-\equiv c_1-c_3$ are related by an algebraic
constraint  (shown as the dashed line in figure \ref{fig:ted})  \cite{nostrong}. 
 A more exotic, but viable, subset of the parameter space can be considered in which $c_1=c_3=0$. The PPN and pulsar 
 constraints do not apply here and a cosmological analysis is potentially the only way of constraining the values of the
 coupling constants. We shall label this alternate space of models as ${\cal E}$. 
 Note that, in what follows, we will write down the equations in a general form and then 
 study the two subspaces ${\cal C}$ and ${\cal E}$ independently.

We now focus on cosmological scales and assume a homogeneous and isotropic background spacetime in which the metric is of the form $g_{ab}dx^{a}dx^{b} =-{dt}^{2}+a(t)^{2}\gamma_{ij}dx^{i}dx^{j}$
where $t$ is physical time, $\gamma_{ij}$ is the identity matrix, and $a(t)$ is the scale factor. Throughout this letter, subscripts $i$ and $j$ will run 1 to 3.
The vector field must respect the spatial homogeneity and isotropy of the system and so will only have a non-vanishing `$t$' component; this constraint
fixes $A^{b} = (1,0,0,0)$. The energy-momentum tensor of the
matter  will include the standard menagerie: photons, neutrinos, baryons, dark matter, and the cosmological constant.

In this background, the t-t component of the aether stress energy tensor is equal to $(3/2)\alpha H^{2}$
where $\alpha\equiv c_1+3c_2+c_3$. For models ${\cal C}$ we have $\alpha=-2c_+c_-/(c_++c_-)$. The fractional energy densities in the various components are given
by $\Omega_{i}(a)\equiv 8\pi G\rho_{i}(a)/3H_{0}^{2}$ and $\Omega_{AE}= (\alpha/2)\Sigma_{i}\Omega_{i}/(1-\alpha/2)$ where 
$H_0=100h$ km s$^{-1}$Mpc$^{-1}$ is the
Hubble constant today and $\rho_i$ is the energy density in the fluid component $i$.
In quasistatic spacetimes the aether exhibits tracking behaviour such that the locally measured value of Newton's constant is actually
$G_{N}=G/(1+c_{14}/2)$ \cite{CL}.  For  models ${\cal C}$ we have $c_{14}=2c_+c_-/(c_++c_-)$ \cite{simplify}. Thus, given a value $\rho_{i}$, the actual $\Omega_{i}$ is related to the value $\Omega_{Ni}$ inferred using a locally measured value of $G_N$ as: $\Omega_{i}=(1+c_{14}/2)\Omega_{Ni}$. Hence, explicitly using our expression for $\Omega_{AE}$, the Friedmann equation becomes:
\begin{eqnarray*}
H^{2}=H_{0}^{2}\frac{2+c_{14}}{2-\alpha}\sum_{i}\Omega_{Ni} \nonumber
\end{eqnarray*}

To fully explore the cosmological consequences of the aether, we must consider
linear perturbations around the background. We will
do so in the synchronous gauge and use conformal time coordinates: given $
g_{ab}dx^{a}dx^{b}=-a^{2}d\tau^{2}+a^{2}[\gamma_{ij}+h_{ij}]dx^{i}dx^{j}$, the two scalar potentials $\eta$ and $h$ are defined by:
$
h_{ij}(\textbf{x},\tau)= d^{3}k e^{i\textbf{k}\cdot\textbf{x}}[\hat{k}_{i}\hat{k}_{j}h(\textbf{k},\tau)+(\hat{k}_{i}\hat{k}_{j}-\frac{1}{3}\delta_{ij})6\eta(\textbf{k},\tau)].
$
The aether field can be written as
$
A^{d}=\frac{1}{a}(1,\partial_{i}V)
$; the zeroth component is fixed equal to $a^{-1}$ by the gauge choice and the fixed-norm constraint.
Instead  of $V$ itself, we choose to to use the variable:
$\xi \equiv V-\frac{1}{2k^{2}}(h+6\eta)' $
with which the evolution equations take a particularly instructive form. 

The gravitational field equations are
\begin{eqnarray*}
(1-\frac{1}{2}\alpha )k^{2}\eta' = 4\pi Ga^{2}i k^{j}\delta T^{0}_{\phantom{0}j}  +\frac{1}{2}k^{4}c_{123}\xi \nonumber
\end{eqnarray*}
and
\begin{eqnarray*}
(1+\frac{1}{2}c_{14})({\cal H}h'-2k^{2}\eta) &=& -8\pi Ga^{2}\delta T^{0}_{\phantom{0}0}   \nonumber\\
\nonumber &&-\frac{1}{2}(c_{14}+\alpha)6{\cal H}\eta' -\frac{3}{2}c_{14}\Sigma_{f} \\
\nonumber && +c_{14}(1+c_{+})k^{2}(\xi'+2{\cal H}\xi).
\end{eqnarray*}
For models ${\cal C}$ we have $c_{123}=2c_+^2/3(c_++c_-)$.

The aether equation of motion is:
\begin{eqnarray*}
0 &=& c_{14}(1+c_{+})\xi''+2{\cal H}c_{14}(1+c_{+})\xi' \nonumber \\
\nonumber &&+ [2c_{14}(1+c_{+})(\frac{a''}{a}-{\cal H}^{2})\\
\nonumber &&-(c_{14}+\alpha)(\frac{a''}{a}-2{\cal H}^{2})+c_{123}k^{2}]\xi \\
\nonumber && +(c_{14}+\alpha)\eta'+(c_{14}+\alpha)\frac{1}{k^{2}}({\cal H}^{2}-\frac{1}{2}\frac{a''}{a})(h'+6\eta')\\
\nonumber && -\frac{3}{2}\frac{c_{14}}{k^{2}}\Sigma_{f}',
\end{eqnarray*}
where ${\cal H}$ is the conformal Hubble parameter, primes are derivatives with respect to $\tau$ and  
$\Sigma_{f} \equiv -8\pi G a^{2} (\hat{k}_{i}\hat{k}^{j}-\frac{1}{3}\delta^{j}_{\phantom{j}i})\Sigma^{i}_{\phantom{i}j}$, 
where $\Sigma^{i}_{\phantom{i}j}$ is the traceless component of the fluid stress energy tensor.
The homogeneous `sourceless' solution to the above equation during an era where $a\propto \tau^{n}$ is
$
\xi(k,\tau) = \tau^{1-2n}[f_1(k)J(\beta,c_{s}k\tau)+f_2(k)Y(\beta,c_{s}k\tau)]
$ where $f_{i}$ are functions to be fixed by boundary conditions, J and Y are Bessel  functions and the various constants are
defined through:
$b_{1} \equiv -2n-({c_{14}+\alpha})(n^{2}+n)/[{c_{14}(1+c_{+})}]$, 
$\beta \equiv  (1-8n+b_{1}+b_{1}^{2})^{1/2} $ and
$c_{s}^{2} \equiv c_{123}/(c_{14}(1+c_{+}))$.
With $c_s^2$ positive, the solutions are damped  and oscillatory solutions when $c_{s}k\tau \gg 1$ and power law  when $c_{s}k\tau \ll 1$ \cite{ZFS2}.

It was shown in \cite{lim} that the primordial scalar power spectrum ${\cal P}_{\Phi}$ (where $\Phi$ is the trace perturbation to the metric in the conformal Newtonian gauge) is modified relative to that of a Universe with no aether, $\tilde{{\cal P}}_{\Phi}$,  through
%\begin{equation}
${\cal P}_{\Phi}=\tilde{{\cal P}}_{\Phi}\left[\frac{1-\frac{\alpha}{c_{14}}c_{+}}{1+c_{+}}\right]^{2}$
%\end{equation}
whilst $\xi$ and $\xi'$ are driven to a vanishingly small value compared to their values at the onset of inflation. We will
work with the equivalent initial conditions in the synchronous gauge.

\begin{figure}[htbp]
\center
\epsfig{file=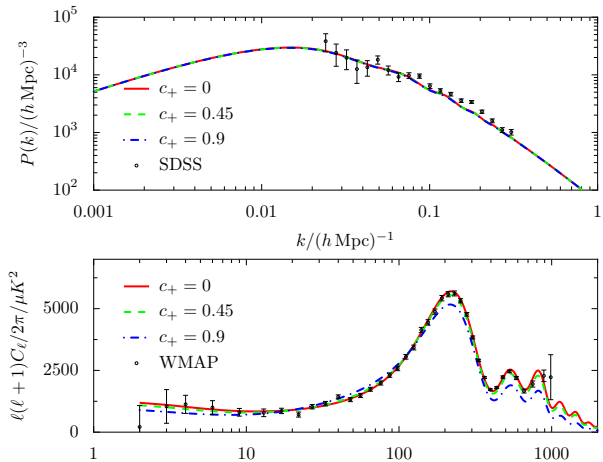,scale=0.67}
\vspace{-15pt}
\caption{The angular power spectrum of the CMB (bottom) and the power spectrum of galaxies (top) for a sample
of class ${\cal C}$ Einstein-Aether models, with different $c_+$  (with $c_-$ chosen to satisfy the weak field binary pulsar constraint --- the dashed line
of figure \ref{fig:ted}) The other parameters have their $\Lambda CDM$ best fit values, with the $\Omega_i$ rescaled as described in the text.
Superposed are the WMAP and SDSS datasets.}
\vspace{-10pt}
\label{fig:C}\end{figure}

To study these effects in detail, we have modified the Boltzmann code \emph{CMBEASY}\cite{cmbeasy} by adding  a Newton-Raphson solver for the Hubble parameter, and including the aether components in the density and pressure; the perturbation evolution has been modified by adding  $\xi$ and $\xi'$ as the integrated components. 
 In Figure \ref{fig:C} we show the effect of the aether on the angular power spectrum of anisotropies in the CMB 
and the power spectrum of galaxies (we superpose the WMAP and SDSS
data) for a selection of parameters in class ${\cal C}$.

The dominant effect for smaller values of $c_+$ is on the large-scale CMB, through the integrated Sachs-Wolfe effect;  it leads to a supression on large scales (which curiously
enough is favoured by large scale CMB data).  As expected, the overall growing mode of matter perturbations is very weakly affected and the change on the power spectrum of galaxies is marginal.

\begin{figure}[htbp]
\center
\epsfig{file=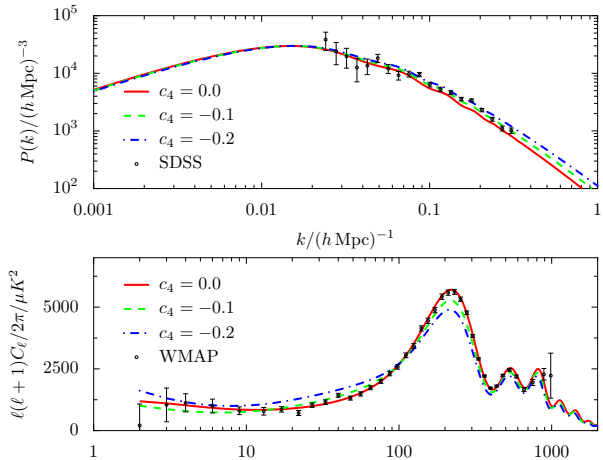,scale=0.67}
\vspace{-15pt}
\caption{The angular power spectrum of the CMB (bottom) and the power spectrum of galaxies (top) for a sample
of exotic class ${\cal E}$ Einstein-Aether models where $c_1=c_2=c_3=0$.  The other parameters have their $\Lambda CDM$ best fit values, with the radiation density modified to account for the change in the gravitational constant.
Superposed
are the WMAP and SDSS datasets.}
\vspace{-10pt}
\label{fig:E}\end{figure}

As is usual in cosmological model testing, we compute parameter constraints using a Monte-Carlo Markov Chain (MCMC)\cite{MCMC}.
We explore a 6 dimensional parameter space consisting of the fractional baryon density, $\Omega_b$, the fractional matter density, $\Omega_M$, the
Hubble constant, $H_0$, the scalar spectral index, $n_S$, the optical depth, $\tau_D$, the overall amplitude of fluctuations, the bias factor of SDSS galaxies and
the two aether parameters, $c_+$ and $c_-$. We constrain parameters using the WMAP 3-year release, the Boomerang 03 release and data from ACBAR and VSA\citep{cmbdata}, as well as the SDSS and $2DF$ surveys\cite{SDSS,2DF}. We also use measurements of the luminosity distance as a function of redshift from supernovae Ia measurements \cite{SN}  but have found
that these data sets have very little ability to constrain this class of models.

The marginalized constraints from the CMB and large-scale structure on the two aethereal parameters in model ${\cal C}$ are shown in figure \ref{fig:ted} \cite{conventions}.   The best-fit aether model is mildly superior to standard $\Lambda CDM$ cosmology, at about $2\sigma$, at a cost of two extra parameters. 

\begin{figure}[htbp]
\epsfig{file=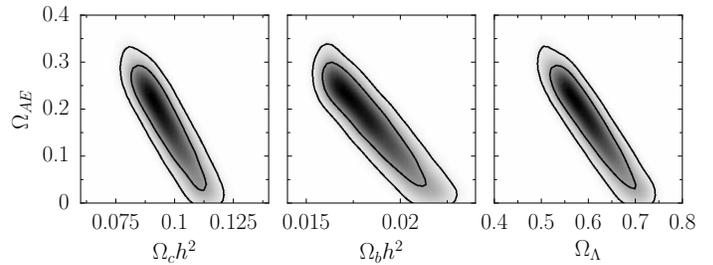,scale=0.5}
\vskip -0.1in
\caption{Joint constraints on the fractional aether density, $\Omega_{AE}$, with the physical dark matter density, $\Omega_c h^2$, 
the physical baryon density, $\Omega_b h^2$ and the fractional $\Lambda$ density, $\Omega_\Lambda$.  
The contours are $1$ and $2\sigma$. }
\label{figure2d}
\vskip -0.2in
\end{figure}

The soft lower limit at $c_- > -0.5$ comes from a prior on the baryon fraction.  This signals an
important characteristic of these models: the strong correlation between the fractional energy density in the aether, $\Omega_{AE}$, and
the other energy components.  This is perhaps the primary result of our analysis and is illustrated in figure \ref{figure2d}: the CMB and LSS data restrict the background to evolve
as in the $\Lambda$CDM case, which in turn leads to a rescaling of the different energy components in the presence of the
aether.  Naturally this also affects the constraints on the other cosmological parameters.  These constraints, under the $\Lambda CDM$ and aether models  with and without the weak binary pulsar constraint, are shown in table \ref{tab:params}.  As expected, the largest shift is seen in the $\Omega_i$.

\begin{figure}[htbp]
\vskip -0.2in
\begin{center}
\epsfig{file=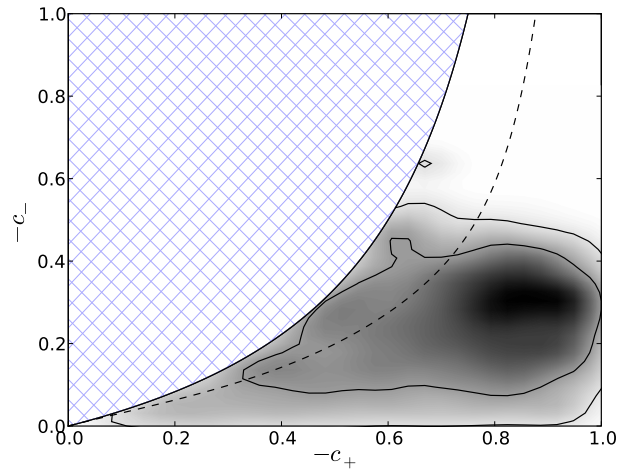,scale=0.45}
\vskip -0.15in
\caption{Joint constraints on the parameters $-c_+$ and $-c_-$.  The black lines are the $1$ and $2\sigma$ contours, where we have marginalized over the values of the parameters.  The hatched region is excluded by \v{C}erenkov constraints;  the dashed line indicates where weak-field constraints from binary pulsars are met.  Both are taken from  \cite{jacobson1}.}
\label{fig:ted}
\end{center}
\vskip -0.2in
\end{figure}

\begin{table}[htdp]
\begin{center}
\begin{tabular}{|c|c|c|c|}
\hline
Parameter & $\Lambda CDM$ & General &  Weak Pulsar \\
\hline
$\Omega_c h^2$ 	 &  $0.137 \pm 0.004 $  	& $ 0.097 \pm 0.01 $ &		$ 0.098 \pm 0.01 $\\
$\Omega_b h^2$  	 & $0.022 \pm 0.001 $  &$  0.019 \pm 0.002 $ & 	$ 0.019 \pm 0.002 $\\
$H_0$ 			 & $69.7 \pm 1.6 $  		&$71.7 \pm 2.0 $ &			$ 72.5 \pm 2.4 $\\
$\tau_D$ 		 	 &$ 0.08 \pm 0.029 $		&$0.077 \pm 0.027 $ &		$ 0.078 \pm 0.028 $\\
$n_s$ 			 & $0.955 \pm 0.015 $ 	&$ 0.976 \pm 0.02 $ &		$ 0.984 \pm 0.024$\\
$\Omega_{\Lambda}$ &$ 0.671 \pm 0.019 $	&$0.61 \pm 0.05 $ &			$ 0.67 \pm 0.028$\\
$c_1$ & 		---						&$ -0.46 \pm 0.14 $&	$-0.26 \pm 0.12 $\\		
$c_2$ & 		---						&$ 0.34 \pm 0.1 $&		$0.20 \pm 0.09 $\\
$c_3$ & 		---						&$ -0.23 \pm 0.1 $&	$-0.12 \pm 0.05$\\
$c_4$ & 		---						&$ 0.13 \pm 0.09 $&		$0.05 \pm 0.02$\\
\hline
\end{tabular}
\caption{Mean and $1\sigma$ error values of the marginalized likelihoods for a range of cosmological
parameters. The left most column is for $\Lambda CDM$ with no aether, the central column for general class ${\cal C}$ models, and the right hand column for class ${\cal C}$ models with the weak pulsar constraint (on the dashed line of Figure 1).}
\end{center}
\vskip -0.3in
\label{tab:params}
\end{table}

As stated above, the CMB and LSS play the dominant role in generating these constraints, and interestingly enough this is through the change in
the background evolution and its effect on the metric perturbations, and not necessarily through the presence of perturbations
in the vector field. Indeed, artificially switcing off the perturbations in the aether field has essentially no effect on the power
spectrum of LSS and a small effect (of approximately $10\%$) on the angular power spectrum of the CMB.

So far we have focused on models in class ${\cal C}$, where we found that the coupling constants
are allowed to vary quite widely. In the case
of models in class ${\cal E}$, the cosmological data are far more restrictive. For example, fixing $c_1=c_2=c_3=0$ we
find that $-0.05 < c_4 <0$ (note that in this case $\Omega_{AE}=0$). If we allow $c_2$ to be non-zero we find that
both $c_2$ and $c_4$ must be in $[-0.01,0]$.     The reason for this constraint is illustrated in figure \ref{fig:E}; at low $\ell$ the integrated Sachs-Wolfe effect induced by the modified potentials is large enough to disrupt the $C_\ell$.  These are the strongest constraints on these parameters that currently exist.

In this letter we have studied the effect of Lorentz violation on cosmology as parametrized by the Einstein-Aether model.  We have found that Lorentz violation in this form is compatible with current cosmological data and,
combined with other non-cosmological probes we have found constraints  on $c_+$ and $c_-$.  The data also require the rescaled combination of density parameters, in which the background evolution is unchanged from a $\Lambda CDM$ universe.  We have also found tight constraints on the other allowed range of parameter space, $\cal{E}$, which has, until now, been relatively unconstrained by
other methods.  Collectively these constraints arise from tests on distance scales spanning more than fifteen orders of magnitude.  

There are of, course,
other possible ways of parameterizing Lorentz violation which are not encompassed by the Einstein-Aether model.  In particular one may relax the
fixed-norm constraint on the aether field \cite{gripaios} or couple it directly to the matter content of the Universe \cite{kostelecky}.
Such theories tend to have a much stronger effect on the evolution of the background cosmology \cite{fgsz} or lead to distinct
experimental signatures \cite{exp}. Hence the results presented here are currently the most comprehensive (though conservative) constraints on the $c_i$s, and thus on Lorentz-violating vector theories.

{\it Acknowledgments}:
We thank T. Jacobson, B. Foster, C. Skordis and G. Starkman for useful discussions.  We are extremely grateful to D. Mota for discussions and for access to computer codes relating to earlier work.
TGZ is supported by an STFC studentship and JAZ by an STFC Rolling Grant.

%\tighten
%\vspace{-.3in}


\begin{thebibliography}{99}
\vspace{-.7in}

%%data
\bibitem{weinberg} S.~Weinberg, ``Quantum Theory of Fields'', CUP, (2005)
\bibitem{QG} G.~Amelino-Camelia, Int J Mod Phys D 11, 35 (2002)
\bibitem{kostelecky} D.~Colladay and V.A.~Kostelecky, Phys Rev D 58, 116002 (1998)
\bibitem{jacobson1} T.~Jacobson, arXiv:0801.1547 (2008)
\bibitem{WN}C.M.~Will, K.~Nordvedt, ApJ, 177, 757 (1972); ApJ, 177, 775 (1972)
\bibitem{jacobson2} D.~Garfinkle, C.~Eling and T.~Jacobson, Phys Rev D 76, 024003 (2007), C.~Eling, T.~Jacobson, Class Quant Grav 23, 5643 (2006)
\bibitem{mattingley} T.~Jacobson, D.~Mattingley, Phys Rev D 70, 024003 (2004)
\bibitem{foster} B.~Foster, T.~Jacobson, Phys Rev D 73, 064015 (2006)
\bibitem{CL} S.M.~Carroll, E.A.~Lim, Phys Rev D, 70, 123525 (2004)
\bibitem{lim}E.A.~Lim, Phys Rev D 71, 063504 (2005)
\bibitem{kanno} S.~Kanno, J. Soda, Phys Rev D 74, 063505, (2006)
\bibitem{ZFS2}T.G.~Zlosnik, P.G.~Ferreira, G.D.~Starkman, Phys Rev.  D 74, 044037, (2006); Phys Rev D 75, 044017, (2007)
\bibitem{H1} A.~Halle, H.~Zhao, B. Li, arXiv:0711.0958
\bibitem{mota}B.~Li, D.~Mota, J.~Barrow, Phys Rev D 77 024032 (2008)
\bibitem{conventions} We use the signature $(-,+,+,+)$ in accord with \cite{bfosterweak}. Hence our coupling coefficients, $c_i$ will have the opposite
signs to those of \cite{jacobson1}
\bibitem{cerenkov} J.W.~Elliott, G.D.~Moore, H.~Stoica, JHEP 0508:066, (2005)
\bibitem{bfosterweak} B.Z.~Foster, Phys Rev D 73,104012 (2006); Erratum-ibid. D 75,129904 (2007)
\bibitem{nostrong}For systems with strong internal fields, 
a number of other effects must be taken into consideration. Though the exact degree to which these effects may further limit the parameter space is still uncertain, 
 it has been argued \cite{jacobson1,bfosterstrong} that they may limit the $c_{i}$ to be of the order $10^{-2}$.  We do not use this constraint in this letter.
\bibitem{bfosterstrong} B.Z.~Foster, Phys Rev D76, 084033 (2007)
\bibitem{simplify} Note that if the PPN constraints are satisfied then the relation $c_{14} = -\alpha$ automatically holds.
\bibitem{cmbeasy} M.~Doran, JCAP 0510:011 (2005)
\bibitem{seljak_zaldarriaga} U.~Seljak, M.~Zaldarriaga, ApJ. 469, 437 (1996)
\bibitem{MCMC} A.~Lewis, S.~Bridle, Phys Rev  D66 103511 (2002); J.~Dunkley et al, MNRAS 356:925 (2005)
\bibitem{cmbdata}J.~Dunkley et al arXiv:0803.0586 (2008); C.~MacTavish et al, ApJ 647,799 (2006); C.~Kuo et al, ApJ, 600,32 (2004); K.~Grainge. et al MNRAS 341 L23 (2003)
\bibitem{SDSS} M.~Tegmark et al ApJ .606,702 (2004)
\bibitem{2DF} S.~Cole et al, MNRAS 362,505 (2005)
\bibitem{SN} A.~Riess et al, ApJ. 607,665 (2004); P.~Astier et al, Astron.Astrophys. 447:31-48, (2006)
\bibitem{gripaios} B.M.~Gripaois, JHEP, 10, 069 (2004)
\bibitem{fgsz}P.G.~Ferreira et al, Phys Rev D 75 044014 (2007)
\bibitem{exp}V.A.~Kostelecky, C.D.~Lane, Phys Rev D 60,116010 (1999).

\end{thebibliography}
\end{document}